\documentstyle[namedreferences,psfig,solphys]{spackapp}
\def\verbfont{\small\tt}

\newcommand{\citeN}[1]{\citeauthor{#1}\ (\citeyear{#1})}
\newcommand{\citeNP}[1]{\citeauthor{#1},\ \citeyear{#1}}

\newcommand{\ujlisti}{
\itemsep=0 em
\parsep=0.5 em
\partopsep=0.25 em
\topsep=0 em}
\newcommand{\ujlistii}{
\itemsep=0 em
\parsep=0.5 em
\partopsep=0.25 em
\topsep=0 cm}

\renewcommand{\[}{\begin{equation}}
\renewcommand{\]}{\end{equation}}
\newcommand{\scri}{\scriptsize} 
\newcommand{\vc}[1]{\mbox{\bf #1}}
\newcommand{\mx}[1]{\hat{#1}}

\newcommand{\tfrac}[2]{{\textstyle\frac{#1}{#2}}}

\def\lang{\langle}
\def\rang{\rangle}

\runningtitle{Evolution of global magnetic field}
\runningauthor{Petrovay \& Szak\'aly}

\begin{opening}

\title{TRANSPORT EFFECTS IN THE EVOLUTION OF THE GLOBAL SOLAR MAGNETIC FIELD}
\author{K. \surname{Petrovay}}
\institute{Instituto de {Astrof\'\i sica} de Canarias, 
	   La Laguna, Tenerife, E-38200 Spain} 
\author{G. {Szak\'aly}}
\institute{E\"otv\"os University, Department of Astronomy, Budapest, 
              Ludovika t\'er 2, H-1083 Hungary} 
\end{opening}

\begin{ao}
\\
K. Petrovay\\
Instituto de Astrof\'\i sica de Canarias\\
La Laguna, Tenerife, E-38200 Spain\\
E-mail: kpetro@iac.es
\end{ao}                   

\begin{document}

\begin{abstract}
The axisymmetric component of the  large-scale solar magnetic fields has a
pronounced poleward branch  at higher latitudes. In order to clarify the origin
of this branch we construct an axisymmetric model of the passive transport
of the mean  poloidal magnetic field in the convective zone, including
meridional  circulation, anisotropic diffusivity, turbulent pumping and density
pumping. For realistic values of the transport coefficients we find that
diffusivity is  prevalent, and the  latitudinal distribution of the field at
the surface simply reflects the  conditions at the bottom of the convective
zone. Pumping effects concentrate the field to the bottom  of the
convective zone; a significant part of this pumping occurs in a shallow
subsurface layer, normally not resolved in dynamo models. The phase delay of
the surface poloidal field relative to the bottom poloidal  field is found to
be small. These results support the double dynamo wave models, may  be
compatible with some form of a  mixed transport scenario, and exclude the
passive transport theory for the origin of the polar branch.

\end{abstract}

\section{Introduction}
The time--latitude distribution of the axisymmetric component of the 
large-scale solar magnetic fields is known to have a pronounced poleward branch 
at higher latitudes (\citeNP{Stenflo:ApSS}, \citeyear{Stenflo:Helsi},
\citeyear{Stenflo:NATO}; \citeNP{Stenflo+Gudel}; \citeNP{Ribes+Bonnefond}; 
\citeNP{Mouradian+Soru-Escaut}). 
This 
branch is also present in the butterfly diagram of a number of tracers of the 
magnetic field such as quiescent filaments, polar faculae, or the coronal green 
line (\citeNP{Callebaut+Makarov}; \citeNP{Makarov+Sivaraman};
\citeNP{Leroy+Noens}). 

Several conflicting explanations exist for this polar branch. Noting that the 
separation latitude of the two branches, 30--40${}^\circ$, approximately 
coincides with the latitude where the radial differential rotation changes sign 
according to helioseismology, one group of 
theories interprets it as the surface reflection of a high-latitude poleward 
propagating dynamo wave, coexisting with the low-latitude equatorward wave
(\citeNP{Gilman+:layer.dyn}; \citeNP{Belvedere+:GAFD}, 
\citeyear{Belvedere+:Nature}). The Parker--Yoshimura 
rule of sign (\citeNP{Belvedere:SPhrev}) then naturally leads to the correct 
directions of propagation if $\alpha$ is negative, as expected in the lower 
overshooting layer where the dynamo should operate. (In fact this is only 
so for certain latitudinal profiles of $\alpha$, cf. \citeNP{Schmitt:Potsdam}.)
The most modern version of 
these models, also incorporating some transport effects, is due to 
\citeN{Rudiger+Brandenburg}. In what follows we will refer to these models as 
\it double wave\/ \rm models.

An alternative approach to the problem of the origin of the poleward drift is 
to interpret it in terms of magnetic transport processes in the solar
photosphere and convective zone. The common ancestor of all such models was the 
now classic Babcock--Leighton interpretation of the solar cycle 
(\citeNP{Babcock:merid.circ}, \citeNP{Leighton:diffusion}). The 
Babcock--Leighton model was what one 
may call a \it mixed transport model\/ \rm in as much as the poloidal fields 
were brought to the surface in a concentrated form in active regions, and 
thereafter they were passively transported to the poles by 
transport processes (diffusion and meridional circulation). In the more 
recent mixed transport models (\citeNP{Wang+:1.5D}; 
\citeNP{Choudhuri+:mixed.transp}) meridional circulation plays the main role in 
transporting the weak fields to the poles, and is also supposed to be 
responsible for the equatorward drift of toroidal fields in the dynamo layer. 
(In the original Babcock--Leighton theory the equatorward branch was still 
supposed to be due to a dynamo wave.)

A third type of models may be called \it passive transport \/\rm models. 
These are a variation of the mixed transport models where the 
``active source'' of the weak fields in the form of active regions is 
also substituted by a passive transport mechanism. Dikpati \& Choudhuri 
(\citeyear{Dikpati+Choudhuri:1}, \citeyear{Dikpati+Choudhuri:2}) suggested such 
a scenario wherein the poloidal fields are brought to the surface at low 
latitudes by meridional circulation. In these models the origin of the 
equatorward drift is not specified; instead, it is just given as a lower 
boundary condition. The resulting butterfly diagrams for the poloidal field 
were in good agreement with the observations provided that the value of the 
turbulent magnetic diffusivity used in the calculations was artifcially reduced 
to about 10 km${}^2$/s, i.e.\ nearly 2 orders of magnitude lower than expected. 

In the present paper we generalize the models of Dikpati \& Choudhuri 
(\citeyear{Dikpati+Choudhuri:1}, \citeyear{Dikpati+Choudhuri:2}) to include all 
turbulent transport effects such as anisotropic turbulent diffusion, turbulent 
pumping, and density pumping, as well as the meridional circulation. We 
construct models with realistic values of the transport coefficients in order 
to see whether the horizontal component of pumping can support the meridional 
circulation in producing a poleward migration despite the strong diffusive
link between the surface and the bottom of the convective zone. (Such a 
possibility was suggested by \citeNP{Krivod+Kichat} and 
\citeNP{Kichat:pump.dynamo}.) Another 
motivation for this study is to check to what extent is the radial pumping able 
to concentrate the fields to the bottom of the convective zone, thereby 
restricting dynamo action to that layer, and to reduce the surface field 
strength to the observed values, as suggested by \citeN{Schussler:vort.pump}, 
\citeN{Petrovay+Szakaly:AA1}, and \citeN{Petrovay:Freibg}. 

\section{The model}
We calculate the mean poloidal magnetic field in the solar convective zone 
as a function of space and time, assuming axial symmetry and antisymmetry to 
the equatorial plane. No dynamo is supposed 
to operate in the convective zone, i.e.\ the $\alpha$-effect is neglected. The 
dynamo layer below is not included in our computational volume; instead, the
processes operating there are assumed to influence our model via the lower 
boundary conditions. 

The equations of passive magnetic field transport with these assumptions are 
(\citeNP{Petrovay:NATO}):
\begin{eqnarray}
   \partial_t A_\phi= &&a_{\theta\theta}\partial^2_\theta A_\phi 
   +a_{\theta r}\partial_\theta\partial_r A_\phi + a_{rr} 
   \partial^2_r A_\phi+ \nonumber \\
   &&a_\theta\partial_\theta A_\phi +a_r\partial_r A_\phi 
   +a_0 A_\phi  \label{eq:poltransp}  \label{eq:main} 
\end{eqnarray}
\[ \lang B_\theta\rang =-A_\phi /r-\partial_r A_\phi \quad 
   \lang B_r\rang =\partial_\theta A_\phi /r +\cot\theta A_\phi /r   \]
where
\begin{eqnarray}
   & a_{\theta\theta}=b_{\theta\theta}= \beta_{\theta\theta}/r^2 \qquad  &
   a_{\theta r}=b_{\theta r}= 2\beta_{\theta r}/r \nonumber \\
   && a_{rr}=b_{rr}=\beta_{rr} 
\end{eqnarray}
\[ a_\theta = (\gamma_\theta +\tilde\gamma_\theta -U_\theta )/r 
   +(\beta_{\theta\theta }\cot\theta -\beta_{\theta r})/r^2    \]
\[ a_r =(\gamma_r+\tilde\gamma_r -U_r) +(\beta_{rr}
  +\beta_{\theta\theta} +\beta_{\theta r}\cot\theta )/r \]
\begin{eqnarray}
  a_0=&&[(\gamma_r +\tilde\gamma_r -U_r) +\cot\theta (\gamma_\theta 
  +\tilde\gamma_\theta -U_\theta )]/r \nonumber \\
  &&-[\beta_{rr}+\beta_{\theta\theta}\cot^2\theta 
  +\beta_{\theta r}(1+\cot\theta )]/r^2 .  
\end{eqnarray}
$\vc B$ is the mean magnetic flux density, $A_\phi$ is the azimuthal component
of  the mean vector potential, $\vc U$ is the velocity of meridional
circulation,  $\mx\beta$ is the anisotropic turbulent diffusivity, and
$\vec\gamma$ and  $\tilde{\vec\gamma}$ are the normal and anomalous components
of the pumping  (incorporating both the density gradient and the gradient of
turbulence  intensity, \citeNP{Moffatt:transport}). Note that in another
widespread notation $\vec\gamma$ and  $\tilde{\vec\gamma}$ are expressed by
components of the antisymmetric and  symmetric parts of the $\alpha$-tensor.
For $\vc U$ we use the same expression  as given in
\citeN{Dikpati+Choudhuri:1}. For the diffusivity and  pumping standard
mean-field theory expressions are used; the formulae are  collected in the
Appendix. For the evaluation of those expressions we use the  UKX convective
zone model (\citeNP{UKX}). The differential rotation is assumed to  be
independent of depth in the computational regime. 

For the solution of Eq.~(\ref{eq:main}) we employ the following boundary 
conditions. At the bottom of the computational regime 
${A_\phi}={A_\phi}(\theta,t)$ was 
explicitly given, as mentioned above. For the other boundaries we set
\begin{eqnarray} 
  &{A_\phi}=0 \qquad &\mbox{at\ } \theta=0
 \label{eq:pole} \\
 &\partial_\theta {A_\phi}\ =0 \qquad &\mbox{at\ } \theta=\pi/2
 \label{eq:equator} \\
 &\partial_r {A_\phi}=-{A_\phi}/r \qquad &\mbox{at the surface} 
 \label{eq:top}
\end{eqnarray} 
The latter condition implies that the field is vertical at the surface 
(\citeNP{Yoshimura:dynamo}; \citeNP{Brandenburg:CUPrev}). Note that these simple 
boundary 
conditions are not necessarily unrealistic: the field is intermittent, and if 
most of the flux in the photosphere is present in flux tubes exceeding 
$10^{18}$\,Mx then Eq.~(\ref{eq:top}) should apply owing to the strong buoyancy 
of these tubes. If, on the other hand, the flux is mainly present in thinner 
tubes, then a potential field boundary condition would be better (cf.\ 
\citeNP{Petrovay:NATO} for a discussion of this problem). We performed some 
test runs with different forms of the boundary condition to find that this 
choice does not have a great influence on the results.

Equation (\ref{eq:main}) was solved by the alternating direction implicit (ADI) 
method on a 64$\times$64 grid. The grid we use is generally uniform in both 
$\theta$ and $r$. However, immediately below the surface the scale heights of 
the density and of the turbulent velocity are very small. A grid uniform in $r$ is 
unable to resolve this fine structure and the corresponding potentially 
important pumping effects. For this reason, in some calculations a non-uniform 
grid is employed, resolving these layers. (In practice this is realized by 
introducing a new variable $\tilde r=c_1\tanh (c_2 r-c_3)$, rewriting 
eq.~(\ref{eq:main}) in $\tilde r$, and solving it on a grid uniform in 
$\tilde r$.) 

   \begin{figure}[htbp]
\centerline{\psfig{figure=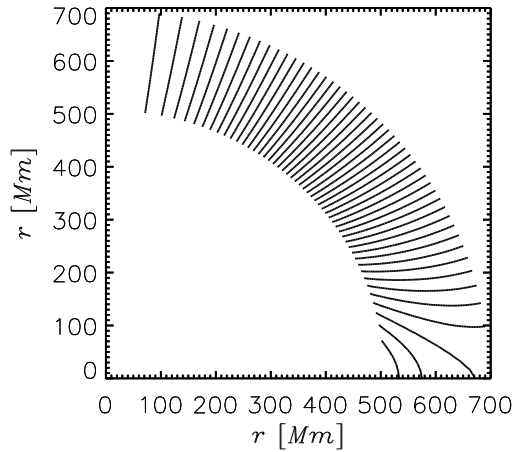,width=9 cm}}
      \caption{Field line configuration in a meridional plane for the 
        stationary dipole model}
         \label{fig:dipcross}
   \end{figure}

\section{Results}
\subsection{Stationary dipole}
\begin{figure}[htbp]
\centerline{\psfig{figure=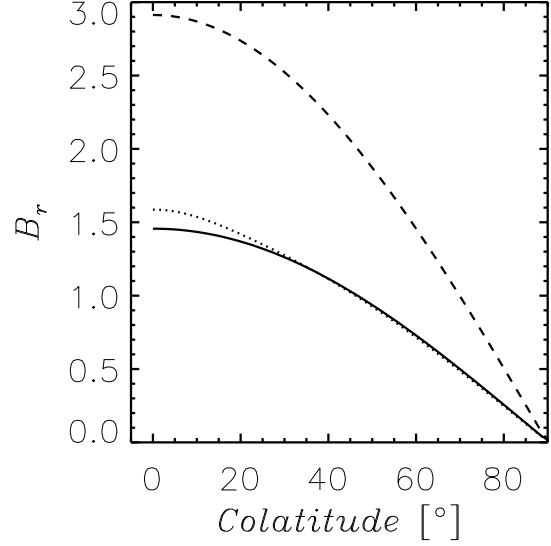,width=9 cm}}
      \caption{Strength of the radial field component as a function of 
        latitude near the surface for the stationary dipole model (solid). 
        Dashed: Same at lower boundary. Dots: Model with rotation and 
        circulation switched off.
        }
        \label{fig:dipBtheta}
   \end{figure}

\begin{figure}[htbp]
\centerline{\psfig{figure=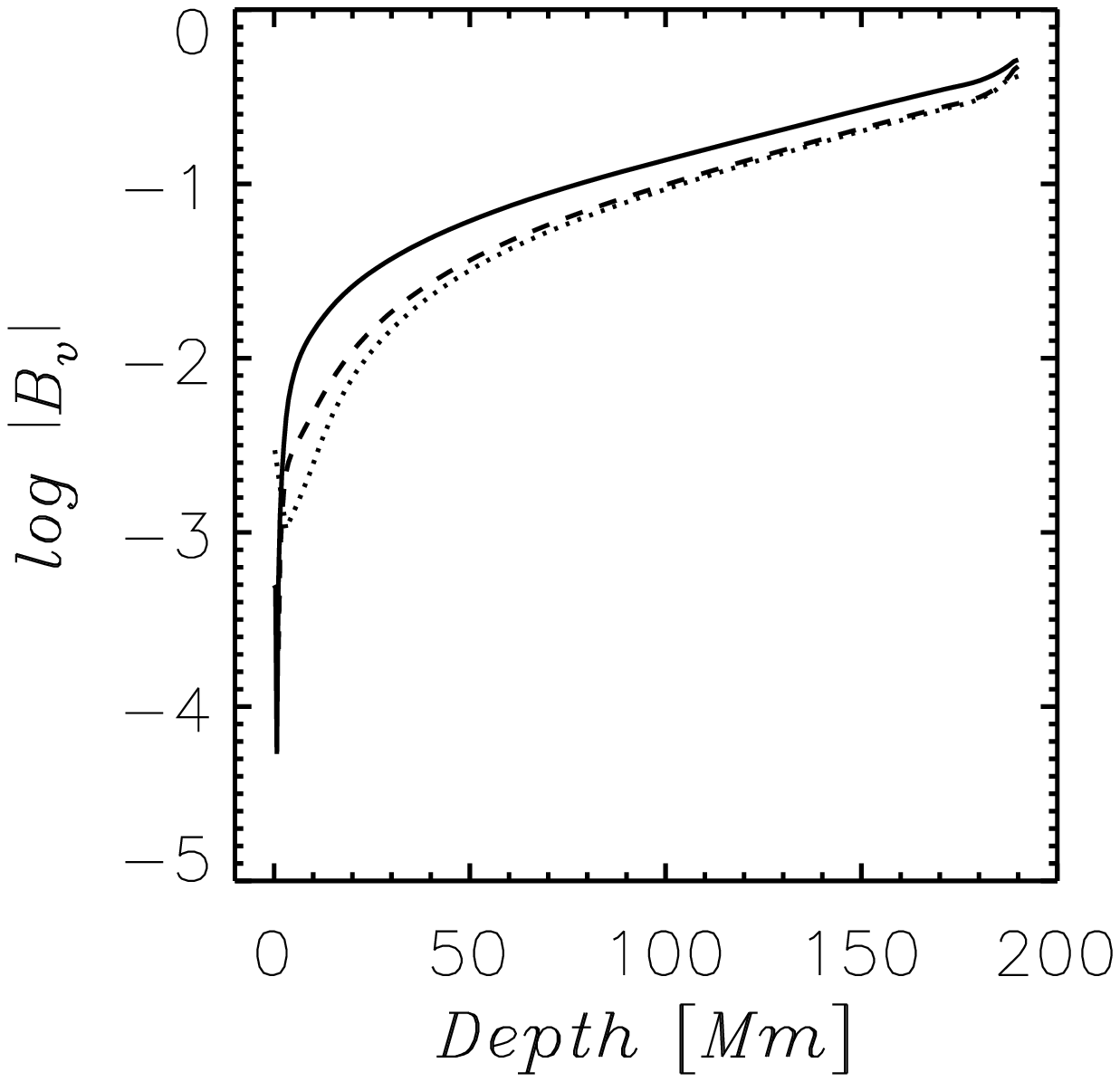,width=9 cm}}
      \caption{Strength of the horizontal field component as a function of 
        depth at $\theta=45^\circ$ for the stationary dipole model (solid). 
        Dots: Same with a uniform grid (shallow layers not resolved). Dashed:
        Model with rotation and circulation switched off.}
         \label{fig:horizfield}
   \end{figure}

It is instructive to consider the case when the lower boundary condition 
corresponds to the vector potential at fixed $r=500\,$Mm of a time-independent 
dipole field centered on the solar center.
The cross section of the resulting field configuration after reaching a 
stationary state is shown in Fig.~\ref{fig:dipcross}. Field lines are nearly 
radial, as a consequence of diffusivity. The effects of rotation and meridional 
circulation are negligible (Fig.~\ref{fig:dipBtheta}). The overall field 
strength varies only by a factor of 2 between the surface and the bottom of the 
convective zone. The horizontal field component, however, increases by three 
orders of magnitude from the surface to the bottom, as a consequence of the 
pumping (Fig.~\ref{fig:horizfield}). This is the only case where the effect of 
rotation is not quite negligible; besides, it is apparent that a proper 
resolution of the shallow layers reduces the surface value by nearly one order 
of magnitude. The overall field structure is however not greatly distorted by a 
uniform grid, neither here, nor in any of the other cases studied.

\subsection{Oscillating dipole}
The case when the lower boundary condition is a dipole with a dipole moment 
oscillating sinusoidally around zero was also considered, in order to see 
whether transport effects can turn this ``standing dynamo wave'' into a 
migrating wave. The result was negative: owing to the overwhelming diffusivity, 
the surface field pattern simply reflects the pattern at the bottom with a 
slight time delay. This finding is contrary to the result of 
\citeN{Kichat:pump.dynamo} who in a simplified model found that the effect of 
pumping, while weak, is sufficient to turn a high-latitude standing wave 
component into a migrating wave. While differences in the details of the models 
may also contribute to the different results, we believe that perhaps the most 
important difference is that in Kichatinov's model the region of flux 
transport spatially coincides with the region of the dynamo (i.e.\ it is a 
convection zone dynamo). 

   \begin{figure}[htbp]
\centerline{\psfig{figure=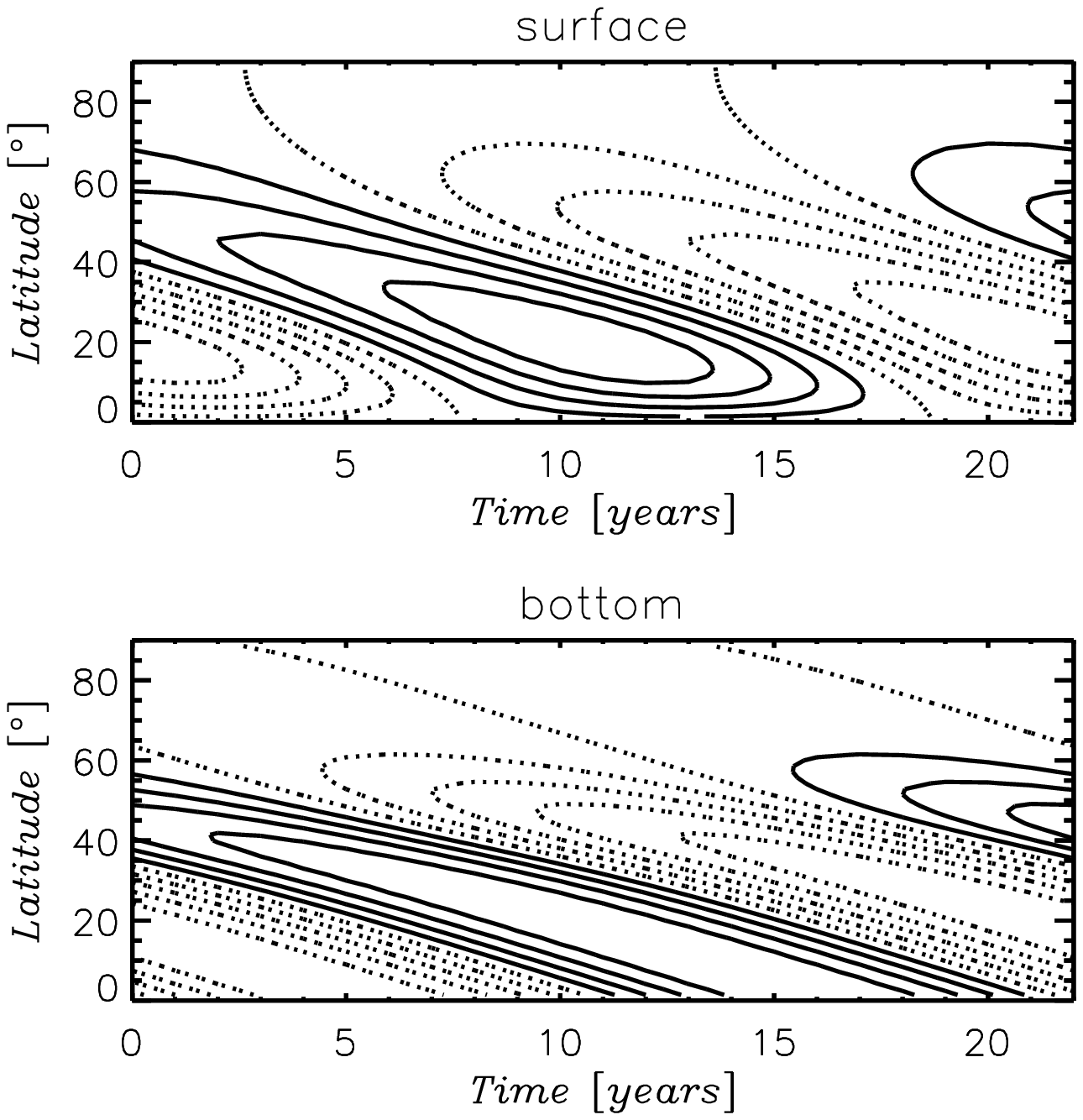,width=9 cm}}
      \caption{Butterfly diagram (i.e.\ $B_r$ vs.\ $\theta$ and $t$) 
      near the surface (\it top\/\rm ) and at the 
      bottom of the convective zone (\it bottom\/\rm ) for a model with an 
      equatorward propagating wave ($\Gamma_1=0$, $\lambda_1=\pi/2$, $A_2=0$).}
         \label{fig:3bfly}
   \end{figure}

   \begin{figure}[htbp]
\centerline{\psfig{figure=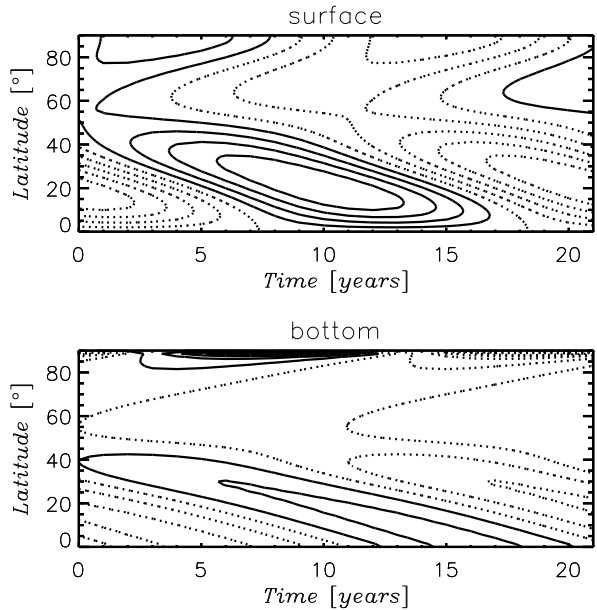,width=9 cm}}
      \caption{Butterfly diagram near the surface (\it top\/\rm ) and at the 
      bottom of the convective zone (\it bottom\/\rm ) for a migrating 
      double wave model ($A_1/A_2=4$, $\Gamma_1=-\Gamma_2=8$,  
      $\lambda_1=-\lambda_2=4\pi/9$, $\delta_1=0$, $\delta_2=\pi/4$).}
         \label{fig:4bfly}
   \end{figure}

\subsection{Migrating field}
The lower boundary condition here was
\[ A_\phi=F(\theta, t; A_1, \Gamma_1, \lambda_1, 0)+
     F(\theta, t; A_2, \Gamma_2, \lambda_2, \delta_2)
\]
where
\begin{eqnarray}
   F(\theta, t; A_i, \Gamma_i, \lambda_i, \delta_i)&&= A_i\left\{1+\exp\left[
    \Gamma_i(\pi/4-\theta)\right]\right\}^{-1}\times \nonumber \\
    &&\cos\left[\omega t+2\pi/\lambda_i
    (\pi/2-\theta)+\delta_i\right]
\end{eqnarray}
with $\omega=9.1\cdot10^{-9}$/s the cycle frequency.

   \begin{figure*}
\centerline{\psfig{figure=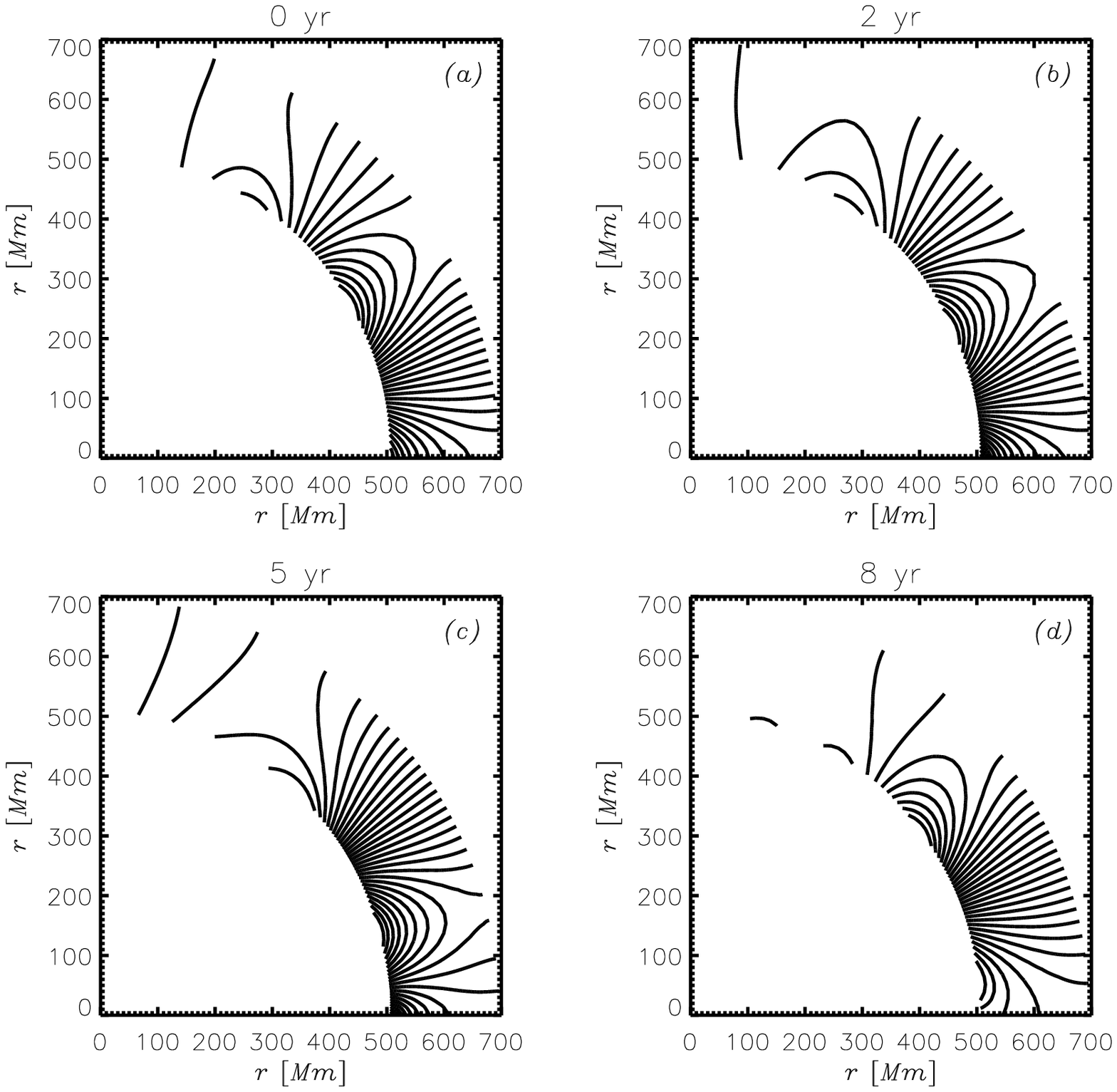,width=13 cm}}
   \vskip -0.8 cm
      \hfill      
      \caption{Field line configurations in a meridional plane 
        at 4 equidistant cycle phases for the migrating double wave model}
        \label{fig:4cross}
    \end{figure*}

\begin{figure}[htbp]
\centerline{\psfig{figure=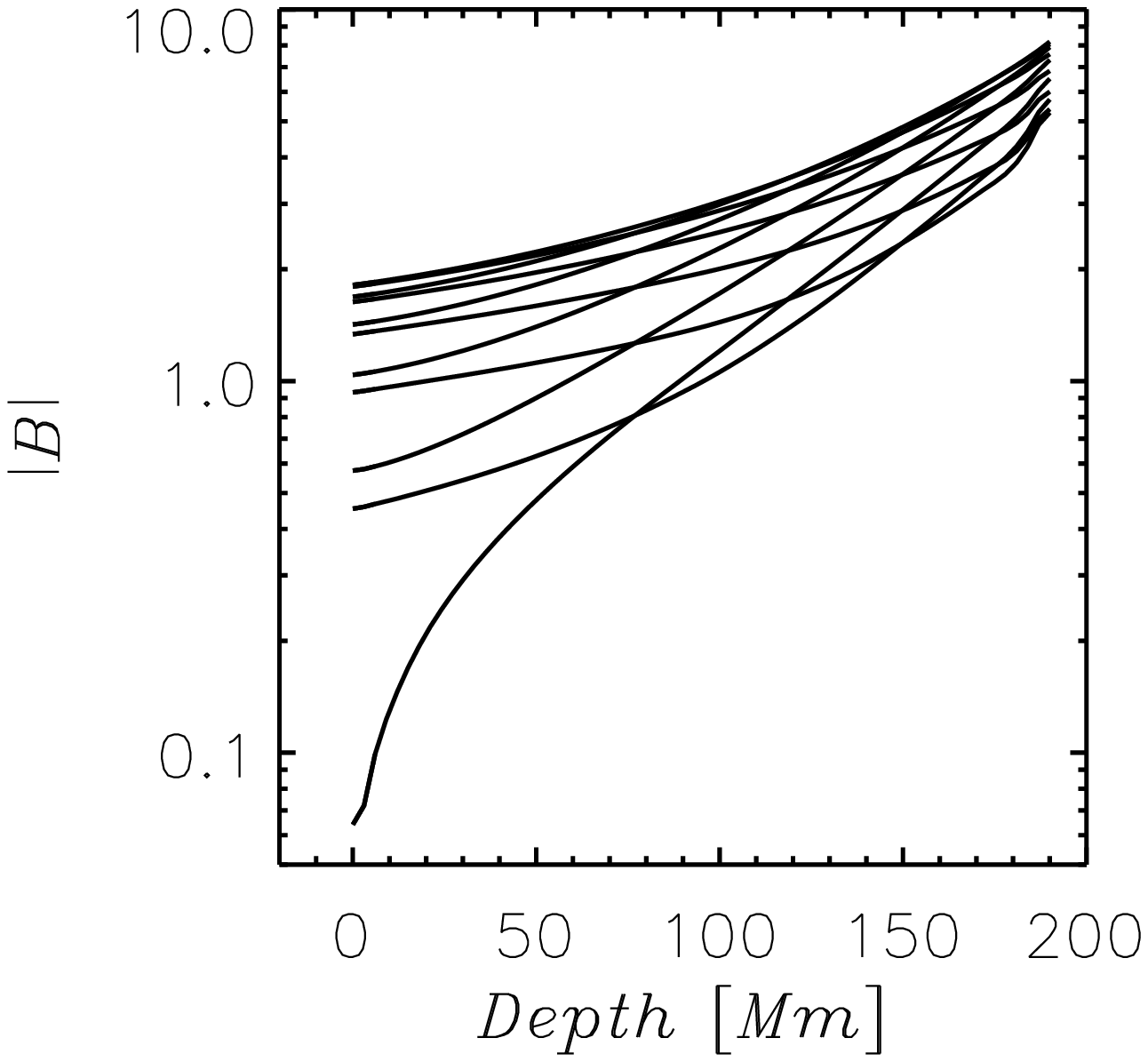,width=9 cm}}
      \caption{Total field strength as a function of depth at $\theta=45^\circ$
        at different cycle phases for a migrating double wave model}
         \label{fig:|B|r}
   \end{figure}

The butterfly diagrams at the surface essentially reflect those at the bottom 
of the convective zone (Figs.~\ref{fig:3bfly} and \ref{fig:4bfly}). A 
high-latitude poleward branch and a low-latitude equatorward branch can thus 
only be produced at the surface if such a pattern is given as input at the 
bottom. Figure~\ref{fig:4cross} shows that the field lines do not show a 
complicated topology inside the convective zone, although they are not exactly 
vertical either as the time dependence does not allow sufficient time for 
diffusion to smooth them. As a consequence, the strong downwards pumping of 
the horizontal field component is to some extent also ``felt'' by the vertical 
component, and, in contrast to the stationary dipole case, the overall field 
strength is also significantly concentrated to 
the bottom of the convective zone (Fig.~\ref{fig:|B|r}). 

\section{Conclusion}
In our axisymmetric model of the passive transport of the large-scale mean 
poloidal magnetic field in the solar convective zone we found that the 
latitudinal distribution of the field at the surface reflects the conditions at 
the bottom of the convective zone, i.e.\ in this regard the convective 
zone behaves as a ``steamy window''. This is due to the fact that with 
realistic values of the transport coefficients diffusivity is prevalent over 
all other effects. Passive transport theories of the origin of the poleward 
branch of the solar butterfly diagram are thus not viable, while double wave 
models are supported by these results. Owing to the large diffusivity the phase 
delay of the surface poloidal field relative to the bottom poloidal field is 
minimal. Coupled with the approximately $\pi$ phase difference between the 
toroidal and poloidal field components for negative $\alpha$ value 
(\citeNP{Dikpati+Choudhuri:2}), this could bring the double wave models to 
accordance with the observed phase relationship.

A mixed transport scenario of the kind 
proposed by \citeN{Wang+:1.5D} is not ruled out (note that our 
Figs.~\ref{fig:dipcross} and \ref{fig:4cross} agree with the field structure 
suggested in that paper), 
but the large diffusivity implies that the photospheric patterns should 
pervade the whole convective zone, being continuously reprocessed through it. 
Note that in fact the active source used in 
those models may also extend to deeper layers of the zone, as emerging magnetic 
loops are thought to shred a significant amount of flux during their rise 
(\citeNP{Petrovay+Szakaly:AA1}; \citeNP{FMI+:explosion}; 
\citeNP{Petrovay+FMI:erosion}).

The vertical distribution of the field (Figs.~\ref{fig:horizfield} and 
\ref{fig:|B|r}) is strongly concentrated to the bottom of the convective zone 
owing to pumping effects. This is in agreement with the earlier proposal of 
\citeN{Schussler:vort.pump} and the one-dimensional results of 
\citeN{Petrovay+Szakaly:AA1} 
and \citeN{Petrovay:Freibg}. The effective pumping may contribute to 
restricting the dynamo to the overshooting layer and it can reduce the surface 
field by about an order of magnitude, thereby offering a solution to the 
overtly high surface poloidal field values found by \citeN{Rudiger+Brandenburg} in 
their double wave model. Note that a significant part of this pumping effect 
occurs in a shallow subsurface layer, normally not resolved in dynamo models.

In summary, our results support the double wave models, may be compatible with 
some form of the mixed transport scenario, and exclude the passive transport 
theory.

In order to decide between the double wave and mixed transport models, two 
key issues should be solved in the future. First, note that in the mixed 
transport model the direction of migration is opposite for poloidal and 
toroidal fields but independent of latitude. Thus, if e.g.\ a clear signature of 
migrating high-latitude toroidal fields were found, this could solve the 
problem in either way, depending on the direction of migration. 
\citeN{Harvey:NATO} claims that high latitude ephemeral active regions show an 
equatorward migration, while \citeN{Callebaut+Makarov} claim that at least 
50\,\% of polar faculae (well known for their poleward drift) correspond to 
dipoles with a preferential east--west orientation, thus forming part of the toroidal 
field. It has even been claimed 
that the highest-latitude part of the sunspot butterfly diagram also shows a 
poleward drift (\citeNP{Becker:poleward}).
A clarification of this issue would clearly be important. 

The role of causality relations between the two branches should also be 
clarified in both theories. Surges in low-latitude activity are followed by 
poleward surges of the high-latitude field, thus supporting the mixed transport 
scenario. On the other hand, the well-known fact that the level of low-latitude 
activity in a sunspot cycle can be predicted from high-latitude fields at 
the end of the previous cycle suggests a causal relationship of the opposite 
sense. It is not impossible that such causal relations can be accommodated in 
both scenarios, but this remains to be seen.
\begin{acknowledgements}
This work was financed in part by the DGES project
no.~95-0028 and by the OTKA under grant no.~F012817.
\end{acknowledgements}

\appendix

\centerline{\bf Expressions for the transport coefficients}

The UKX model assumed fixed anisotropy $\beta_{rr}/\beta_{\theta\theta}=2$
for the diffusivity tensor. We assume that the effect of rotation on the 
diffusivity remains the same as for quasi-isotropic turbulence 
(\citeNP{Kichat:diff.mx}):
\begin{eqnarray}
   &\beta_{\theta\theta}=3\beta(\phi_3 +\phi_2\sin^2\theta) &\quad 
   \beta_{\phi\phi}=3\beta\phi_3 \nonumber \\
   & \beta_{rr}=3\beta (2\phi_3 +\phi_2\cos^2\theta ) &\quad \beta_{\theta r}=
   -\tfrac 32 \beta\phi_2\sin (2\theta) \nonumber \\
   && \quad\beta_{\theta\phi}=\beta_{\phi r}=0 
\end{eqnarray}
\begin{eqnarray}
   &&\phi_2(\Omega_\ast)= \frac 1{4\Omega_\ast^2}
   \left(\frac{\Omega_\ast^2 +3}{\Omega_\ast}\arctan\Omega_\ast -3\right) \\
   && \phi_3(\Omega_\ast)= \frac 1{4\Omega_\ast^2}
   \left(\frac{\Omega_\ast^2 -1}{\Omega_\ast}\arctan\Omega_\ast +1\right)  
\end{eqnarray}
\[ \beta =\tfrac 12 x^2\tau   \]
\[ \Omega_\ast =\tau\Omega /\pi \]
\[ \tau =\mbox{Max}\,\{ H_P/x ; 1000\,\mbox{s}\} \]
where $x$ is the r.m.s.\ radial component of the turbulent velocity in the UKX 
model, $H_P$ is the pressure scale height, and $\Omega=\Omega(\theta)$ is the 
surface plasma differential rotation law. Note that the expression of $\beta$ used 
is valid for non-cellular flows only (free random walk). The cellular nature of the 
supergranular flow on the solar surface may reduce the diffusivity there 
(\citeNP{Ruzmaikin+:cellular}), but even this reduced value is well in excess of
$100\,$km$^2$/s and the reduction should only be confined to the photosphere, as a similar
cellular flow is not expected  in the deeper layers.

The general expressions of the pumping velocities for weak magnetic fields are
\begin{eqnarray}
   \gamma_\theta +\tilde\gamma_\theta = 
   &&\left(K_{\tilde\rho}\tfrac{{}_1{\cal A}^3_2-1}{{}_1{\cal 
   A}^3_2+1}+K_\rho\right) \beta_{\theta r} d_r\rho /\rho \nonumber \\
   &&+K_v\tfrac{{}_1{\cal 
   A}^3_2}{{}_1{\cal A}^3_2+1}(\partial_\theta\beta_{\theta\theta} 
   +2\beta_{\theta r})/r +K_v\partial_r \beta_{\theta r} 
\end{eqnarray}
\begin{eqnarray}
   \gamma_r +\tilde\gamma_r = &&\left( K_{\tilde\rho}
   \tfrac{{}_3{\cal A}^1_2-1}{{}_3{\cal 
   A}^1_2+1}-K_\rho\right) \beta_{rr} d_r\rho /\rho +K_v\tfrac{{}_3{\cal 
   A}^1_2}{{}_3{\cal A}^1_2+1}\partial_r \beta_{rr} \nonumber \\
   &&+K_v\left(\partial_\theta
   \beta_{\theta r} +\beta_{rr} -\beta_{\theta\theta}\right)/r \nonumber \\   
   &&+ K_{\mbox{\scri im}}\tfrac{{}_3{\cal A}^1_2}{{}_3{\cal A}^1_2+1}x 
\end{eqnarray}
\begin{eqnarray}
   &&{}_3{\cal A}^1_2= \left[ 1+(\phi_2/\phi_3)\sin^2\theta \right]^{1/2} \\
   &&{}_1{\cal A}^3_2= 1+\left[ 1+(\phi_2/\phi_3)\sin^2\theta\right]^{-1/2}
\end{eqnarray}
where $K_v=0.6$, $K_\rho=0.15$, $K_{\tilde\rho}=0.55$ and 
$K_{\mbox{\scri im}}=0.1$ 
account for (small) higher-order and intermittency 
corrections (\citeNP{Petrovay:NATO}, \citeNP{Petrovay+Zsargo}); these 
corrections are of minor importance. Note that the magnetic field 
strength does not exceed about 10 G at any point in our computational volume 
(if we assume that the surface fields are about 1 G as observed); so buoyancy 
effects were altogether neglected here. 


\begin{thebibliography}{}

\bibitem[\protect\citeauthoryear{Babcock}{1961}]{Babcock:merid.circ}
Babcock, H.~W.: 1961,
\newblock {\em \apj\/} {\bf 133}, 572

\bibitem[\protect\citeauthoryear{Becker}{1959}]{Becker:poleward}
Becker, U.: 1959,
\newblock {\em Z. f{\"u}r Astrophys.\/} {\bf 48}, 88

\bibitem[\protect\citeauthoryear{Belvedere}{1985}]{Belvedere:SPhrev}
Belvedere, G.: 1985,
\newblock {\em Solar Phys.\/} {\bf 100}, 363

\bibitem[\protect\citeauthoryear{Belvedere, Pidatella, and
  Proctor}{1990}]{Belvedere+:GAFD}
Belvedere, G., Pidatella, R., and Proctor, M. R.~E.: 1990,
\newblock {\em \GAFD\/} {\bf 51}, 263

\bibitem[\protect\citeauthoryear{Belvedere, Proctor, and
  Lanzafame}{1991}]{Belvedere+:Nature}
Belvedere, G., Proctor, M. R.~E., and Lanzafame, G.: 1991,
\newblock {\em Nature\/} {\bf 350}, 481

\bibitem[\protect\citeauthoryear{Brandenburg}{1994}]{Brandenburg:CUPrev}
Brandenburg, A.: 1994,
\newblock in M.~R.~E. Proctor and A.~D. Gilbert (eds.), {\em Lectures on Solar
  and Planetary Dynamos\/}, Cambridge UP, p.~117

\bibitem[\protect\citeauthoryear{Callebaut and
  Makarov}{1992}]{Callebaut+Makarov}
Callebaut, D.~K. and Makarov, V.~I.: 1992,
\newblock {\em Solar Phys.\/} {\bf 141}, 381

\bibitem[\protect\citeauthoryear{Choudhuri, Sch{\"u}ssler, and
  Dikpati}{1995}]{Choudhuri+:mixed.transp}
Choudhuri, A.~R., Sch{\"u}ssler, M., and Dikpati, M.: 1995,
\newblock {\em \aap\/} {\bf 303}, L29

\bibitem[\protect\citeauthoryear{Dikpati and
  Choudhuri}{1994}]{Dikpati+Choudhuri:1}
Dikpati, M. and Choudhuri, A.~R.: 1994,
\newblock {\em \aap\/} {\bf 291}, 975

\bibitem[\protect\citeauthoryear{Dikpati and
  Choudhuri}{1995}]{Dikpati+Choudhuri:2}
Dikpati, M. and Choudhuri, A.~R.: 1995,
\newblock {\em Solar Phys.\/} {\bf 161}, 9

\bibitem[\protect\citeauthoryear{Gilman, Morrow, and {De
  Luca}}{1989}]{Gilman+:layer.dyn}
Gilman, P.~A., Morrow, C.~A., and {De Luca}, E.~E.: 1989,
\newblock {\em \apj\/} {\bf 338}, 528

\bibitem[\protect\citeauthoryear{Harvey}{1994}]{Harvey:NATO}
Harvey, K.~L.: 1994,
\newblock in R.~J. Rutten and C.~J. Schrijver (eds.), {\em Solar Surface
  Magnetism\/}, NATO ASI Series, Kluwer, Dordrecht, p.~347

\bibitem[\protect\citeauthoryear{Kichatinov}{1988}]{Kichat:diff.mx}
Kichatinov, L.~L.: 1988,
\newblock {\em Astron.\ Nachr.\/} {\bf 309}, 197

\bibitem[\protect\citeauthoryear{Kichatinov}{1993}]{Kichat:pump.dynamo}
Kichatinov, L.~L.: 1993,
\newblock in F. Krause, K.-H. R{\"a}dler, and G. R{\"u}diger (eds.), {\em The
  Cosmic Dynamo\/}, IAU symp.~157, Kluwer, Dordrecht, p.~13

\bibitem[\protect\citeauthoryear{Krivodubskij and
  Kichatinov}{1991}]{Krivod+Kichat}
Krivodubskij, V.~N. and Kichatinov, L.~L.: 1991,
\newblock in I. Tuominen, D. Moss, and G. R{\"u}diger (eds.), {\em The Sun and
  Cool Stars: Activity, Magnetism, Dynamos\/}, Proc.\ IAU Coll.\ 130, Springer,
  Berlin, p.~190

\bibitem[\protect\citeauthoryear{Leighton}{1964}]{Leighton:diffusion}
Leighton, R.~B.: 1964,
\newblock {\em \apj\/} {\bf 140}, 1547

\bibitem[\protect\citeauthoryear{Leroy and Noens}{1983}]{Leroy+Noens}
Leroy, J.-L. and Noens, J.-C.: 1983,
\newblock {\em \aap\/} {\bf 120}, L1

\bibitem[\protect\citeauthoryear{Makarov and
  Sivaraman}{1989}]{Makarov+Sivaraman}
Makarov, V.~I. and Sivaraman, K.~R.: 1989,
\newblock {\em Solar Phys.\/} {\bf 123}, 367

\bibitem[\protect\citeauthoryear{Moffatt}{1983}]{Moffatt:transport}
Moffatt, H.~K.: 1983,
\newblock {\em Rep.\ Prog.\ Phys.\/} {\bf 46}, 621

\bibitem[\protect\citeauthoryear{{Moreno-Insertis}, Caligari, and
  {Sch{\"u}ssler}}{1995}]{FMI+:explosion}
{Moreno-Insertis}, F., Caligari, P., and {Sch{\"u}ssler}, M.: 1995,
\newblock {\em \apj\/} {\bf 452}, 894

\bibitem[\protect\citeauthoryear{Mouradian and
  Soru-Escaut}{1991}]{Mouradian+Soru-Escaut}
Mouradian, Z. and Soru-Escaut, I.: 1991,
\newblock {\em \aap\/} {\bf 251}, 649

\bibitem[\protect\citeauthoryear{Petrovay}{1994a}]{Petrovay:Freibg}
Petrovay, K.: 1994a,
\newblock in M. Sch{\"u}ssler and W. Schmidt (eds.), {\em Solar Magnetic
  Fields\/}, Proc.\ Freiburg Internat.\ Conf., Cambridge UP, p.~146

\bibitem[\protect\citeauthoryear{Petrovay}{1994b}]{Petrovay:NATO}
Petrovay, K.: 1994b,
\newblock in R.~J. Rutten and C.~J. Schrijver (eds.), {\em Solar Surface
  Magnetism\/}, NATO ASI Series C433, Kluwer, Dordrecht, p.~415

\bibitem[\protect\citeauthoryear{Petrovay and
  {Moreno-Insertis}}{1997}]{Petrovay+FMI:erosion}
Petrovay, K. and {Moreno-Insertis}, F.: 1997,
\newblock {\em \apj\/} {\bf 485}, 398

\bibitem[\protect\citeauthoryear{Petrovay and
  {Szak\'aly}}{1993}]{Petrovay+Szakaly:AA1}
Petrovay, K. and {Szak\'aly}, G.: 1993,
\newblock {\em \aap\/} {\bf 274}, 543

\bibitem[\protect\citeauthoryear{Petrovay and
  Zsarg{\'o}}{1998}]{Petrovay+Zsargo}
Petrovay, K. and Zsarg{\'o}, J.: 1998,
\newblock {\em \mnras\/} {\bf 296}, 245

\bibitem[\protect\citeauthoryear{Ribes and Bonnefond}{1990}]{Ribes+Bonnefond}
Ribes, E. and Bonnefond, F.: 1990,
\newblock {\em \GAFD\/} {\bf 55}, 241

\bibitem[\protect\citeauthoryear{R{\"u}diger and
  Brandenburg}{1995}]{Rudiger+Brandenburg}
R{\"u}diger, G. and Brandenburg, A.: 1995,
\newblock {\em \aap\/} {\bf 296}, 557

\bibitem[\protect\citeauthoryear{Ruzmaikin and
  Molchanov}{1997}]{Ruzmaikin+:cellular}
Ruzmaikin, A.~A. and Molchanov, S.~A.: 1997,
\newblock {\em {Solar Phys.}\/} {\bf 173}, 223

\bibitem[\protect\citeauthoryear{Schmitt}{1993}]{Schmitt:Potsdam}
Schmitt, D.: 1993,
\newblock in F. Krause, K.-H. R{\"a}dler, and G. R{\"u}diger (eds.), {\em The
  Cosmic Dynamo\/}, IAU symp.~157, Kluwer, Dordrecht, p.~1

\bibitem[\protect\citeauthoryear{Sch{\"u}ssler}{1984}]{Schussler:vort.pump}
Sch{\"u}ssler, M.: 1984,
\newblock in T.~D. Guyenne and J.~J. Hunt (eds.), {\em The Hydromagnetics of
  the Sun\/}, Proc.\ 4th Eur.\ Meeting on \solphys, ESA, p.~67

\bibitem[\protect\citeauthoryear{Stenflo}{1988}]{Stenflo:ApSS}
Stenflo, J.~O.: 1988,
\newblock {\em Ap\&SS\/} {\bf 144}, 321

\bibitem[\protect\citeauthoryear{Stenflo}{1991}]{Stenflo:Helsi}
Stenflo, J.~O.: 1991,
\newblock in I. Tuominen, D. Moss, and G. R{\"u}diger (eds.), {\em The Sun and
  Cool Stars: Activity, Magnetism, Dynamos\/}, Proc.\ IAU Coll.\ 130, Springer,
  Berlin, p.~193

\bibitem[\protect\citeauthoryear{Stenflo}{1994}]{Stenflo:NATO}
Stenflo, J.~O.: 1994,
\newblock in R.~J. Rutten and C.~J. Schrijver (eds.), {\em Solar Surface
  Magnetism\/}, NATO ASI Series C433, Kluwer, Dordrecht, p.~365

\bibitem[\protect\citeauthoryear{Stenflo and G{\"u}del}{1988}]{Stenflo+Gudel}
Stenflo, J.~O. and G{\"u}del, M.: 1988,
\newblock {\em \aap\/} {\bf 191}, 137

\bibitem[\protect\citeauthoryear{Unno, Kondo, and Xiong}{1985}]{UKX}
Unno, W., Kondo, M., and Xiong, D.-R.: 1985,
\newblock {\em \pasj\/} {\bf 37}, 235

\bibitem[\protect\citeauthoryear{Wang, Sheeley, and Nash}{1991}]{Wang+:1.5D}
Wang, Y.-M., Sheeley, N.~R., and Nash, A.~G.: 1991,
\newblock {\em \apj\/} {\bf 383}, 431

\bibitem[\protect\citeauthoryear{Yoshimura}{1975}]{Yoshimura:dynamo}
Yoshimura, H.: 1975,
\newblock {\em \apj Suppl.\/} {\bf 29}, 467

\end{thebibliography}

\end{document}